# A sustainable waste-to-protein system to maximise waste resource utilisation for developing food- and feed-grade protein solutions.

Ellen Piercy [a1], Willy Verstraete [b1], Peter R. Ellis [c1], Johan Rockström [d,¬]-, Pete Smith [e, ¬], Oliver Witard [f¬], Jason Hallett [g,¬] Christer Hogstrand [h, ¬], -Geoffrey Knott [i], Ai Karwati [j], Henintso Felamboahangy Rasoarahona [k], Andrew Leslie [a], Yiying He [a], Mason Banks [a], Miao Guo [a, *]

a. Department of Engineering, Faculty of Natural, Mathematical & Engineering Sciences, King's College London, Strand Campus, London, WC2R 2LS, UK
b. Center for Microbial Ecology and Technology, Ghent University, Coupure Links 653, 9000 Ghent, Belgium and Avecom, Industrieweg 122P, 9000 Ghent, Belgium
c. Biopolymer's Group, Departments of Biochemistry and Nutritional Sciences, Faculty of Life Sciences & Medicine, King's College London, Franklin-Wilkins Building, London, SE1 9NH, UK
d. Potsdam Institute for Climate Impact Research, University of Potsdam, 14412 Potsdam, Germany
e. Institute of Biological and Environmental Sciences, School of Biological Sciences, University of Aberdeen, Aberdeen, AB24 3UU, Scotland, UK
f. Centre for Human and Applied Physiological Sciences, School of Basic and Medical Biosciences, Faculty of Life Sciences and Medicine, King's College London, London, SE1 1UL, UK
g. Department of Chemical Engineering, Imperial College London, South Kensington Campus, London, SW7 2AZ, UK
h. Department of Nutritional Sciences, Faculty of Life Sciences & Medicine, King's College London, Franklin-Wilkins Building, London, SE1 9NH, UK
i. New Foods Ltd, 22 Uxbridge Road, London W5 2RJ, UK
j. Noveltindo Eiyo Tech Ltd, IPB Techno Science Park, Bogor 16128, Indonesia
k. MIKASA, Academic Network for Nutrition, Antananarivo 101, Madagascar

1 equivalent contributions as co-first authors

¬equivalent contributions

* corresponding author: miao.guo@kcl.ac.uk

**Abstract**

A waste-to-protein system that integrates a range of waste-to-protein upgrading technologies has the potential to converge innovations on zero-waste and protein security to ensure a sustainable protein future. We present a global overview of food-safe and feed-safe waste resource potential and technologies to sort and transform such waste streams with compositional quality characteristics into food-grade or feed-grade protein. The identified streams are rich in carbon and nutrients and absent of pathogens and hazardous-contaminants, including food waste streams, lignocellulosic waste from agricultural residues and forestry, and contaminant-free waste from the food and drink industry. A wide range of chemical, physical, and biological treatments can be applied to extract nutrients and convert waste-carbon to fermentable sugars or other platform chemicals for subsequent conversion to protein. Our quantitative analyses suggest that the waste-to-protein system has the potential to maximise recovery of various low-value resources and catalyse the transformative solutions toward a sustainable protein future. However, novel protein regulation processes remain expensive and resource intensive in many countries, with protracted timelines for approval. This poses a significant barrier to market expansion, despite accelerated research and development in waste-to-protein technologies and novel protein sources. Thus, the waste-to-protein system is an important initiative to promote metabolic health across the lifespan and tackle the global hunger crisis.

Despite continuous efforts to achieve the goal of 'zero hunger' Sustainable Development Goals (SDG), the global undernourished population is projected to increase from 688 million to 841 million by 2030 (*1*). A major contributor to this forecast is the occurrence of war and disruptive political situations, and failures to distribute economically accessible food to the poorest societies on our planet. In addition, increasing strains on food security are exacerbated by the unsustainable reliance on finite natural capital resources such as land and water, that are required for traditional farming techniques. Animal-sourced protein is a highly resource-intensive and nutritionally inefficient method of food production based on nitrogen utilisation yet constitutes 18% of the current global protein supply (*2-4*). Indeed, the projected increase in demand for meat protein (to almost double by 2050) poses significant environmental concerns, particularly in relation to land and water availability and greenhouse gas emissions (*5-8*). The Covid-19 pandemic has threatened global food supply chains at multiple levels, causing interruptions to the planting, harvesting, and transportation of crops (*9-11*). Such interruptions exacerbate the issue of food security with the worst post-pandemic scenario estimated to produce 909 million people with undernutrition by 2030 (*12-14*), highlighting the need for a secure yet sustainable food production system.

Rising food waste presents as an abundant resource for alternative protein solutions. It is estimated that one-third of food produced globally is underutilised for reasons related to logistics of supply and demand. This trend is evident in both developed regions with overnutrition and less developed countries with increasing rates of undernutrition, and is equivalent to 1.3 billion tonnes of wasted food which provides sufficient resources to feed 2 billion people worldwide (*15*). Globally, considerable amounts of carbon-containing and nutrient-rich waste are generated from food and drink sector. For instance, in the UK, 1.5 million tonnes of waste is created from the production of meat, dairy, fruits, vegetables and starch products, beverages and brewing, and other food products (*16, 17*).

This review focuses on the contaminant-free organic component of three broad waste streams that can be converted to food-grade or animal feed-grade protein through sustainable protein production technologies. We consider (i) food waste streams present in organic fraction of municipal solid waste (OFMSW); (ii) lignocellulosic waste, which is defined here as the lignocellulosic agricultural residues from crop cultivation (e.g. straw) as well as forestry waste (e.g. wood chips); and (iii) food industry waste in the form of organic gas, liquid, and solid streams generated from processing and manufacturing within the food and drink sector. These waste streams offer considerable potential for resource recovery and protein production due to the high concentrations of nutrients, degradable organic compounds and the absence of pathogens, toxic metals, and other hazardous contaminants.

A range of sustainable technologies are available to extract or convert nutrients and organic compounds present in contaminant-free waste to produce food- or feed-grade protein. Utilisation of microbial biotechnologies such as fermentation can achieve yields of approximately 40% cell biomass from dry waste matter (*18*). At least 80 species have been reported to produce microbial protein, but a better understanding of the microbes involved and their potential for protein recovery from waste is needed (*19*). Higher organisms such as insects can also be used as bio-converters within a waste-to-protein system. These higher organisms typically attain a maximum upgrading efficiency of only 10% but can also yield biomass components of significant functional value. Additionally, biochemical and physical treatments can be used to recover extra nutrients from waste streams, upgrade waste-to-protein systems, or convert waste-carbon to fermentable sugar and other platform chemicals for subsequent conversion to protein. Despite the advances in individual technologies, critical gaps remain in the development of innovative systems that integrate these technologies for optimised protein recovery from diverse waste streams.

In this review, we define a 'waste-to-protein system' as a collection of pathways using process technologies to recover food-grade and/or feed-grade protein from contamination-free organic waste resources. Accordingly, 'waste-to-protein' refers to the proteins derived or produced from non-contaminated food-safe or feed-safe organic materials exhibiting compositional quality suitable for valuable upgrading. Food-grade and feed-grade proteins have differing requirements with regards to feedstock quality (food-safe vs. animal feed-safe, respectively), and must comply with hygienic quality and safety standards set by regulators which vary significantly by country (*20*). The primary aim of this article is to provide an overview of the strategies and pathways with the potential to transform globally abundant contaminant-free waste into a sustainable 'waste-to-protein system' to achieve global protein security and contribute to a circular-economy aspiration (*21, 22*).

The objective of this review is three-fold. Firstly, this article critically evaluates the viability of food-safe and feed-safe waste streams as 'waste-to-protein' resources, with an emphasis on their abundance and biochemical composition. We then appraise the technologies available for waste-to-protein conversion, focusing on three promising, evidence-based pathways: biochemical & physical treatment, microbial protein, and insects as bio-converters. Finally, we propose a sustainable 'waste-to-protein' system that maximises waste resource utilisation for the development of food-grade and feed-grade protein solutions to promote global food security and ameliorate the hunger pandemic.

## Waste-to-protein sources

### Feed-grade organic fraction of municipal solid waste

Annual global household waste generation is equivalent to 2.01 billion tonnes of municipal solid waste (MSW). The organic fraction of municipal solid waste (OMSFW) accounts for around 40% of global MSW generated each year, presenting as an abundant source of feed-grade organic waste for a waste-to-protein system (*23, 24*). It is an overly abundant resource for high-income countries, and a valuable nutrient resource for low-income countries due to its macronutrient profile (*25*). Fig. 1. illustrates the rate of MSW generation by country, as well as the regional composition. Rates of generation range from 4.94 kg/capita/day (Antigua & Barbuda) to 0.14 kg/capita/day (Nepal). While higher quantities of MSW are produced by high-income countries (Fig.1a), low-income countries tend to generate a larger organic fraction (food and garden waste) compared to high-income nations (Fig.1b). On average, 184 g of OFMSW is generated per capita per day with crude protein content ranging from 4.35 g/capita/day (South Asia) to 31 g/capita/day (Caribbean). MSW is projected to increase by 70% in developing countries, and a marked increase in MSW generation has been observed in areas with rapid urbanisation (*15*, *26*). Developing regions such as Africa and South East Asia also account for 91.8% of worldwide undernourishment, highlighting the urgent need to explore new protein solutions, e.g. waste-to-protein, to meet increasing nutrient and protein demands in these areas (*1*).

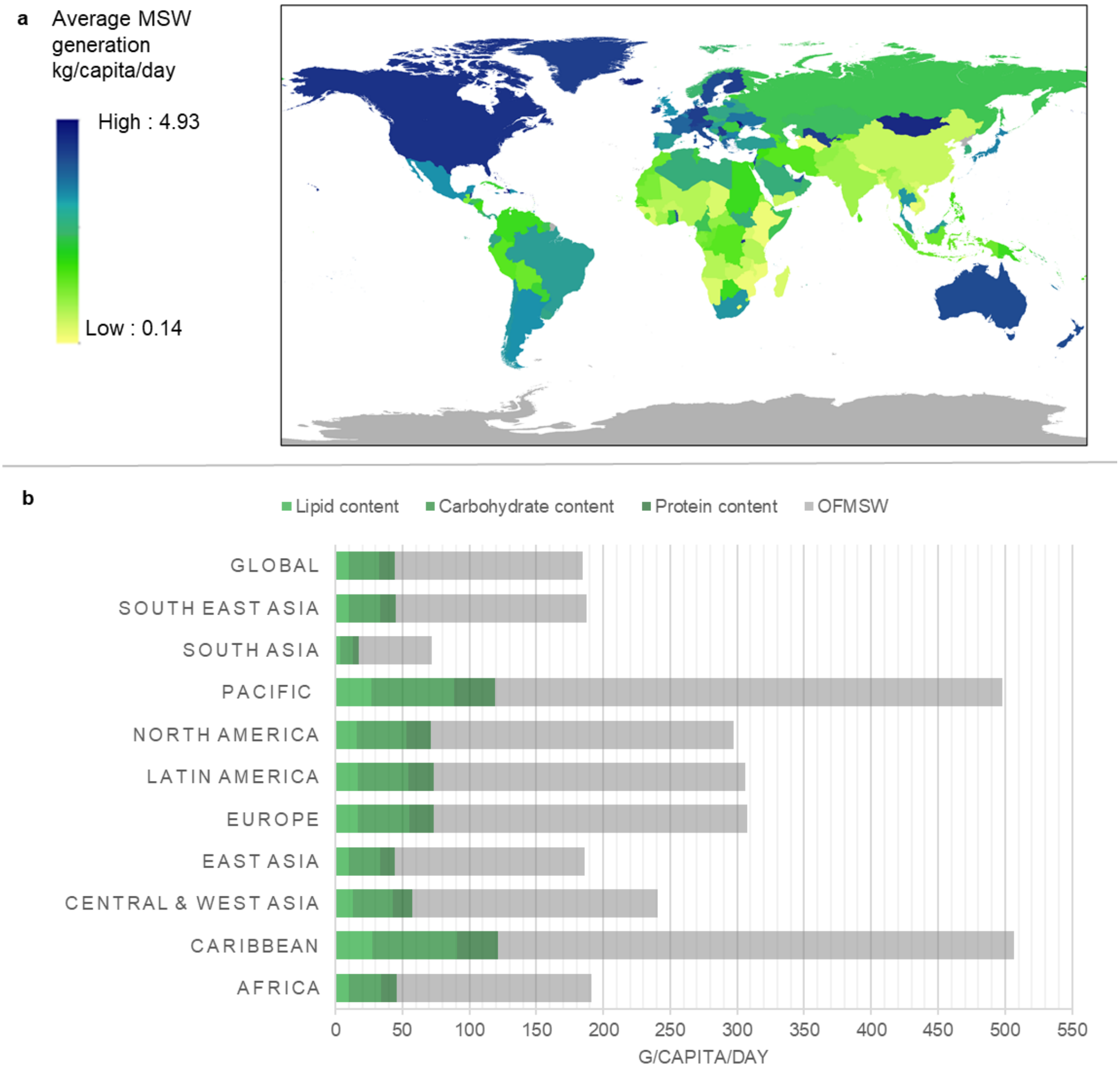

**Fig. 1. Global production of Municipal Solid Waste (MSW). (a)** Average MSW generation (kg/capita/day) was calculated for each country using data from literature (*15, 25, 27, 28*) where MSW generation was plotted according to a colour gradient scale ranging from low (minimum 0.14kg/capita/day) to high (maximum 4.93kg/capita/day). (**b)** Regional OFMSW composition and average lipid, carbohydrate and protein contents (g/capita/day) were calculated from previously reported values (*15, 29*). Detailed data can be found in Supplementary Information SI-1, Supplementary Table ST1.

## Agriculture and Forestry Lignocellulosic Waste

Lignocellulosic waste from agriculture is a globally distributed, carbon-rich, non-contaminated and food-safe resource, presenting as a potential candidate for the recovery of nutritionally valuable protein (*11*). Although different countries and regions exhibit varying production rates of agricultural crops, all countries generate lignocellulosic waste in the form of agricultural residues (*30, 31*). In this review, we define agricultural crops as terrestrial plants cultivated on a large scale including cereal grains, fruits, vegetables, oil crops, and sugar crops. We assessed the potential carbon and nutritional values of food-grade lignocellulosic wastes from agriculture sector by examining the biochemical composition of agricultural residues (Fig.2).

Crude protein content often constitutes less than 8% of agricultural residues. However, sustainable technologies could be deployed to convert the lignocellulosic component to protein. For example, microbial strains capable of metabolising lignocellulosic feedstock could be used to produce food-grade or feed-grade protein. Fig.2a presents the lignocellulosic contents of the main agricultural product residues, ranging from 34% to 60% for lignin, 15% to 43% for cellulose and 17% to 36% for hemicelluloses. We focus on cellulose, hemicelluloses and lignin but acknowledge that other cell wall components (e.g. pectins) and intracellular components (e.g. oligosaccharides and starch) warrant future exploratory research.

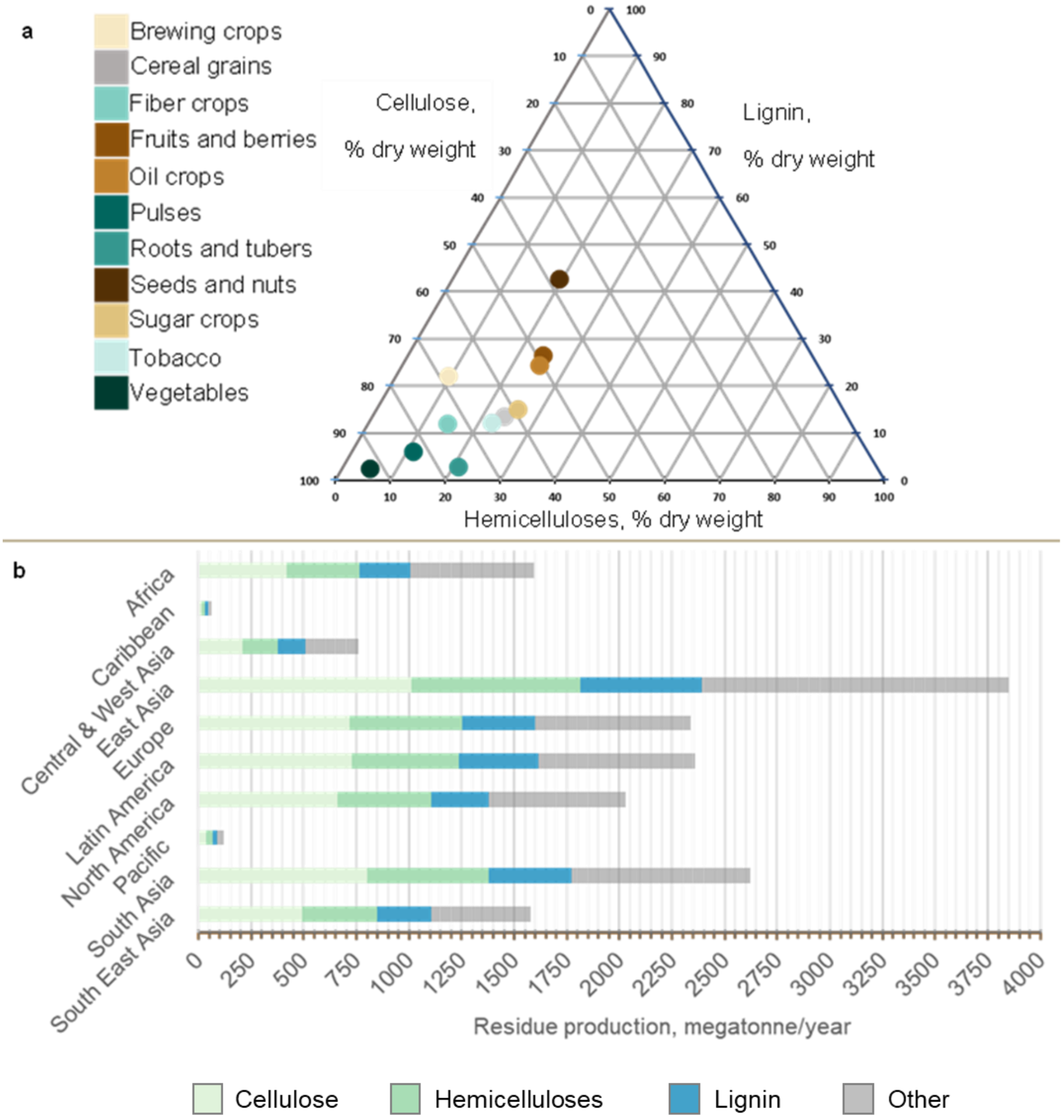

**Fig. 2. Biochemical analysis of agricultural lignocellulosic residues.** Agricultural products were categorised as: brewing crops; cereal grains; fibre crops; fodder; fruits and berries; oil crops; pulses; roots and tubers; seeds and nuts; sugar crops; tobacco; and vegetables. **(a)** Biochemical composition of lignocellulosic component of agricultural product residues based on the Phyllis database (*32*). Values are given as a % of dry weight. **(b)** Regional lignocellulosic production rate and its biochemical composition as part of the total agricultural residue production. Residue production was estimated by applying residue production ratios to production values for 2018 for each region (*32, 33*). Detailed data can be found in Supplementary Information SI-2, Supplementary Table ST2.

Geographical variations in climate and soil conditions contribute to regional differences in production rate and biochemical composition of agricultural residues. East Asia is the largest global producer of lignocellulosic agricultural residues (2,389 megatonnes per year), which constitutes approximately 62% of the total residue production. In comparison, the Caribbean agricultural sector generates only 44 megatonnes per year of lignocellulosic residues, constituting 68% of its total residue production. Overall, total residue production is higher in South and Southeast Asia. However, other regions including both high income and low-income countries also show abundant agricultural residue production, highlighting global potential for lignocellulosic conversion of crop residues to protein (Fig.2b).

Forestry residue is another lignocellulosic waste source (*30, 34, 35*). Global forest resources amount to 600,066 megatonnes/year and comprise of above and below-ground biomass, plus 67,000 megatonnes/year of deadwood. The global distribution and analyses of forestry biomass and corresponding residue biomass can be found in published databases (*36*). Residues generated by forest management, harvesting and processing (particularly in regions with active forestry industries such as Canada and parts of Latin America, and from areas employing tree-cutting for wildfire prevention) could provide substantial lignocellulosic feedstock for a waste-to-protein process system (*37, 38*). The fact that upgrading of lignocellulosic content from forestry residues to human food or animal feed has not taken any dimensions of scale relates to aspects of logistics and particularly cost competitiveness of the products. Furthermore, protein derivation from forestry waste for human consumption is particularly problematic, as forestry land can have significant contamination e.g. those used for phytoremediation.

## Food and Drink Industry Waste

Diverse waste streams are generated from the global food and drink sector, the quantity and composition of which are dependent on the process technologies employed as well as the region of production. This review focuses on two carbon and protein-rich waste streams from the fishing and brewing industries. The shrimp fishing industry is a good target for waste-to-protein resource recovery, being well-established in Africa and South East Asia and generating 6-8 megatonnes/year of protein-rich organic waste (40% protein) during the processing phase (*39*). Shrimp waste also contains chitin, which constitutes 20-30% of its biomass. Chitin can be converted to water-soluble chitosan, a value-added polysaccharide with a range of functional properties and industrial applications (e.g. drug delivery, food thickening and stabilising) (*40, 41*). Combined recovery of protein and value-added polysaccharides such as chitosan has the potential to improve the economics and sustainability of waste-to-protein system processes.

The most abundant by-product generated by the brewing industry is brewer's spent grain (BSG), which offers great potential for protein recovery due to its protein and carbon-rich chemical composition (*42*). The major component of BSG tissues are the cell walls consisting primarily of non-starch polysaccharides (NSP), some of which are lignified (*43*). The NSP include cellulose and non-cellulosic polysaccharides ('hemicelluloses'), particularly arabinoxylans which constitute 25-52% of BSG composition. BSG also has high protein contents, comprising 15-31% of its composition (*44, 45*). Research efforts have focussed on existing chemical processes (e.g. solvent pre-treatment followed by enzymatic hydrolysis) to fractionate the protein components and convert NSP to fermentable sugars for microbial protein production (*46, 47*). However, optimised routes to integration of BSG into the conventional feed and food supply chains using novel processing methods remains as an outstanding research gap.

## Sustainable Protein Production Technologies

Promising technologies presenting sustainable methods of protein recovery include: i) biochemical, chemical, and physical treatments, ii) microbial protein, and iii) insects as bio-converters.

## Biochemical, Chemical and Physical Treatments

A wide range of biochemical, chemical, or physical treatments can be applied to contaminant-free organic waste streams to extract valuable nutrients, amino acids, and proteins, or to transform carbon to fermentable sugar as feedstock for other protein production technologies (*48, 49*).

Membrane filtration (e.g. ultrafiltration, reverse osmosis) and precipitation (e.g. isoelectric precipitation, salting out, organic solvent methods) and adsorption technologies offer great advantages as cost-effective techniques for continuous protein separation.

Membrane filtration has been well-established as a physical treatment to mitigate nutrient concentration and carbon oxygen demand (COD) of industrial effluents, as in the dairy industry to recover value-added caseins and whey proteins from wastewater (*50*). Such methods have demonstrated high efficiency, for example Das et al. (2015) were able to achieve 90% protein recovery from whey waste using combined ultrafiltration and nanofiltration (*51*). Filtration methods are also low in energy consumption and protein denaturement but are challenged by performance issues such as membrane fouling caused by particle deposition and coagulation of charged proteins at the membrane surface. This issue has been observed in various studies, including tuna and dairy wastewater processing, as well as commercially, for example during production trials of flavour enhancer Mycoscent (Quorn), a concentrate containing glutamate and ribonucleotides from mycoprotein wastewater (*52, 53*).

A variety of methods exist to precipitate proteins from solution, including isoelectric precipitation, salting out, and organic solvent methods. Typically, precipitation is a rapid, easily scalable process that can be operated at low temperatures, enabling high throughput, low heat duty and recovery of proteins without denaturement effects. Taskila *et al.* (2017) investigated the use of low-temperature evaporation followed by ethanol precipitation to recover value-added proteins from potato fruit juice. Implementation at pilot scale demonstrated a 50% recovery of proteins from industrial starch waste streams (*54*). Xu *et al.* (2019) studied epigallocatechin-3-gallate (an ester derived from green tea) as a precipitating agent for protein valorisation from soy whey wastewater, achieving a high recovery of 60.7% with a protein purity of 69.51% (*55*). Adsorption technologies have been explored primarily to extract valuable enzymes from waste, as detailed in a review by Shahid *et al.* (2021). Typically, various structural forms of mesoporous silica with modified surface properties are employed for targeted enzyme valorisation and are capable of operating at low temperatures. However, residence time, adsorption capacity and operating pH vary significantly as a function of adsorbent, substrate, and target enzyme of study (*52*).

Despite promising results of new precipitation and adsorption technologies, further studies are required to determine recovery performance when targeting proteins of high nutritional value from a wider range of waste streams. Research efforts focused on adsorbent regeneration/precipitant recovery and recycle capacity are also essential to ensure sustainability and economic viability of the process.

With regards to lignocellulosic waste, fractionation pre-treatment include chemical (e.g. alkali, acid, ionic liquid), thermal (e.g. steam), biological (e.g. ligninolytic microbes) and physical (e.g. extrusion) methods individually or in combination. Extensive research has focussed on pre-treatment technologies, as detailed in several reviews (*56-61*). In short, these reviews conclude that the chemical processes successfully render effective fractionation but introduce design challenges such as solvent recycling and the need for reactor anti-corrosion steps. Physical and thermal routes may lead to cost-effective, solvent-free but energy-intensive solutions. Despite the advantages of low-energy demand and effective lignin depolymerisation, biological routes might be challenged by low reaction rate and inhibitor generation issues.

At the lignocellulosic conversion stage, enzymatic and acid hydrolysis are the most widely adopted technologies to derive fermentable monosaccharides from long-chain carbohydrates. This stage is critical, as the quality of hydrolysate produced significantly impacts downstream processes. In contrast to acid catalysts, enzymes are effective at low temperatures, reducing reactor capital cost and heat duty. However, enzymes remain expensive due to their production complexity, presenting a significant economic barrier to commercial implementation of enzymatic hydrolysis (*56*). Research into lignocellulosic hydrolysis catalysed by acids and enzymes have been detailed in previous reviews (*2, 56, 60, 62*). In brief, according to Modenbach and Nokes (2013) cellulases and xylanases are the most commonly adopted enzymes to degrade cellulose and xylan, respectively (*60*). The degradation mechanisms of these enzymes on their corresponding carbohydrate substrates are discussed by Aditiya *et al.*, (2016). In addition to the common sugars (e.g. sucrose, glucose, fructose, galactose, mannose, ribose, xylose, and arabinose) which occur in nature in the free form, or as monomers of oligosaccharides and polysaccharides, other rare monosaccharides and sugar alcohols (e.g. xylitol, mannitol, erythritol as sugar substitutes) can also be produced by enzyme-catalysed reactions (*61*). The wide range of platform chemicals, in particular the fermentable sugars, provide substrates for the production of microbial protein or alternative protein sources. The capacity of microbial protein produced from such resources to replace conventional protein from animal husbandry was estimated by Pikaar and colleagues (*18*). The authors calculated that in terms of amino acid requirements, up to 10-19% of current global feed crops (occupying 6% of global arable area and equivalent to the entire current cropland of China) could be replaced by microbial protein, freeing up arable land area for other important agricultural practices.

The variety of technology options available offers great potential for novel protein solutions capable of transforming global food production as we know it. For example, Indonesia primarily relies on imported feed-protein such as soybean meal, fish meal and meat bone meal from America and Brazil, exposing the country to feed shortages in the event of global supply chain disruptions (*63*). Recognising this, Indonesian researchers have focused on protein recovery from local palm and coconut oil waste using microbial enzymes (*64*). Transitioning to local waste-to-protein solutions has the potential to significantly improve protein security and sustainability, while reducing the cost of meeting regional and national nutritional demands.

## Microbial protein

Microbial protein technology utilises yeast, fungal, bacterial, or algal strains capable of converting sources of carbon, nitrogen, and oxygen into protein-rich biomass fit for human consumption or animal feeding.

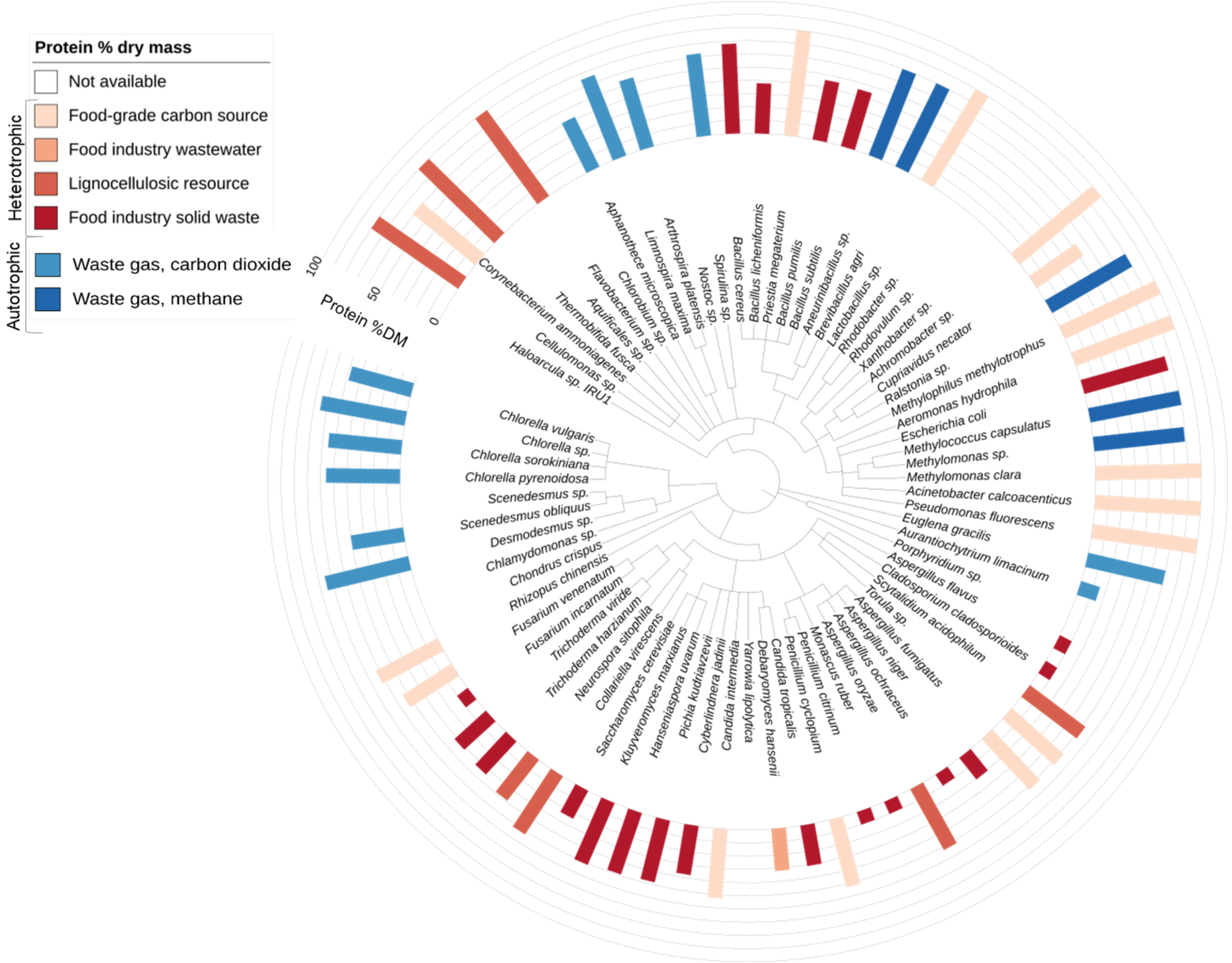

**Fig. 3. Taxonomic tree of reported microbial protein producing species.** Species are sorted according to the National Centre for Biotechnology Information (NCBI) taxonomy database (*65*). Species are grouped by domain: Archaea, Eukaryota or Bacteria. Reported protein contents (% dry mass) are indicated by bar chart ranging from 10% to 80% dry mass (Supplementary Table ST3). Where multiple protein values have been reported an average was calculated. Food-grade carbon source refers to pure food-grade soluble compounds such as glucose, lactose and maltose. Detailed data can be found in the Supplementary Information SI-3 and Supplementary Table ST3.

Approximately 80 different microbial strains have been reported to enable the production of food-grade or feed-grade protein (Fig.3). Microalgae and bacteria represent the most protein-rich sources, within the range of 60-70 wt% and 50-80 wt%, respectively, whereas fungi/yeasts contain approximately 30-50 wt% protein, followed by protists at 10-20 wt% (*66*). The high protein content positions bacteria as a desirable candidate for microbial protein conversion. However, reported palatability issues are yet to be addressed, posing a challenge to the successful commercialisation of bacterial protein as a food product. (*67, 68*)

Fungi have a longstanding history of use in the production of microbial protein food products, some of which are now mass-produced and widely distributed e.g. tempeh. Oncom, a traditional food closely related to tempeh and consumed mainly in West Java, Indonesia, is produced by fermenting *Rhizopus oligosporus* and *Neurospora sitophila*. Interestingly, waste by-products from food production such as soy pulp, peanut press cake and cassava tailings are typically employed as substrates for the fermentation process. Despite serving as a historical waste-to-protein proof of concept, a high quality, mass-produced oncom product has not yet been realised, and very few research efforts have been made to this end (*69*).

As early as the 1970s, a variety of high-quality upgrade products that are rich in microbial protein were established on farm and industrial scales, e.g. volatile fatty acids from Candida yeast (*70, 71*) and methanol to Pruteen (*72*). Despite relative ease of operation and access to a large body of expertise built by long-established fermentation industries, established supply chains (e.g. soybean-based protein) held an economically competitive edge, stifling many early businesses.

Mycoprotein has become one of the most successful food-grade microbial proteins and was originally produced in response to concerns regarding the insufficiency of meat as a sustainable and healthy protein source. It has been commercialised since 1985 as Quorn™ (*73*) and is currently sold in 17 countries, predominantly in Europe but also in developing nations such as the Philippines, and is the largest microbial protein meat alternative (by sales) with over 6 billion meals supplied globally in 2020 (*74*). Quorn™ mycoprotein is produced via fermentation of fungus species *Fusarium venenatum A3/5* utilising glucose as feedstock, with the addition of oxygen, nitrogen, vitamins, and minerals (*75*). Mycoprotein has a moderate protein content (45% of biomass) and contains all essential amino acids (44% of total protein) (*76*). Additionally, it offers positive health attributes compared with animal protein, such as a favourable fatty acid profile and high fibre content (*77*). These properties make Quorn™ mycoprotein well-suited to regions with high prevalence rates of obesity-related diseases such as North America and Europe (*1, 78*). A series of recent studies in human physiology by Monteyne *et al.*, (2020) have examined the capacity for mycoprotein to regulate skeletal muscle protein metabolism in young and older adults, with encouraging results (*79*).

Industrial pioneers have utilised microbial protein technologies to produce protein for human consumption, as well as for animal and aquaculture feed purposes. Notable feed-grade protein products that have been commercialised include All-G Rich® (Alltech), UniProtein® (Unibio) and Feedkind® (Calysta) (*80-82*). Fungal species *Neurospora sitophila* also has a longstanding history of involvement in food production (*83*). White Dog Labs, Inc. (New Castle, Delaware) actively produces microbial protein for animal feed but has not disclosed strain information. Moreover, the carbon transformation company Kiverdi, Inc. (Pleasanton, California) recently introduced 'Air Protein', which converts $CO_2$ to food-grade protein by microbial fermentation, however no detailed information has been disclosed on the hydrogenotrophic microorganisms used (*84*). Solar Foods (Helsinki, Finland) is also using 'air protein' technology to produce food-grade microbial protein (Solein®) via $CO_2$ fermentation at pilot scale and have recently been awarded €35 million in funding. Avecom (Ghent, Belgium) aims to integrate their microbial protein technology with existing food processing businesses as a waste recovery solution, allowing them to produce proteins for food or feed purposes. Furthermore, Avecom's 'Power-to-Protein' research partnership has been investigating renewable hydrogen and atmospheric carbon dioxide as drivers for autotrophic and mixotrophic upgrading of nitrogen from waste to produce feed protein (*85*), however issues of poor hydrogen mass transfer are still being addressed to ensure adequate rates of production. Phototrophic bacteria are also being explored to produce human food and animal feed from secondary resources.

Many microorganisms are still at the research and development stage. Microbial protein production that utilises lignocellulosic waste resources have generated increasing research attention. Two potential technology solutions have been reported, namely *Fusarium venenatum A3/5* fed on lignocellulosic glucose and xylose (*11*) and cellulose-consuming strains such as *Aspergillus niger*, *Neurospore sitaphila*, and *Trichoderma viridae* (*86, 87*). Recently, SylPro® Arbiom has gained attention for scaling up trials of protein production based on the conversion of lignocellulosic forestry waste by yeast species for aquaculture feed (*66, 88*).

Producing novel food ingredients with desirable techno-functional and sensory qualities for use in the food and drink industry remains a formidable challenge (*89*), and the development of microbial protein ingredients is no exception (*85*). Currently, the preferred strategy is to focus on the nutritional value (amino acid composition) of microbial proteins and then search for smart combinations with other food ingredients to provide properties such as taste, texture and structure to the final food, such is the case with current mycoprotein products (*90*). Although there is a large list of potential upgraders, the legislator formulates strict requirements regarding which organisms are accepted as human food. In the European union, applications for novel food status require preparation of detailed technical dossiers as evidence for the safety of products. When added to the considerable costs and complexity involved in the application procedure, this creates an significant barrier to the development and commercialisation of novel foods (*91*) in the EU and in countries adopting a similarly 'cautionary' regulatory model.

**Insects as Bio-converters**

Contamination-free biowaste provides a theoretical feed stream for insects to act as waste-to-protein bio-converters. High conversion rates for *Orthoptera* sp. (1.7 kg feed:1 kg liveweight) (*40*) and *Hermentia illucens* i.e. black soldier fly larvae (1.95-13.42% carbon and 5.4-18.93% nitrogen recycling) have been reported (*92*). The cultivated insects can be harvested and converted into human food through relatively simple processing methods. For example, caterpillars and mealworms are prepared by scalding, drying and cooking (i.e. roasting or boiling), and insect protein bars are prepared by milling and processing (i.e. baking) (*17*, *93*). According to recent estimates, one billion of the world's population are estimated to rely on insects as a primary protein source, particularly in African and South East Asian countries (*94*). Insect-based foods are seeing increasing global acceptance in recent years, with the combined insect market of the US, Belgium, France, UK, the Netherlands, China, Thailand, Vietnam, Brazil and Mexico, predicted to increase from £25 million in 2015 to £398 million in 2023 (*95*).

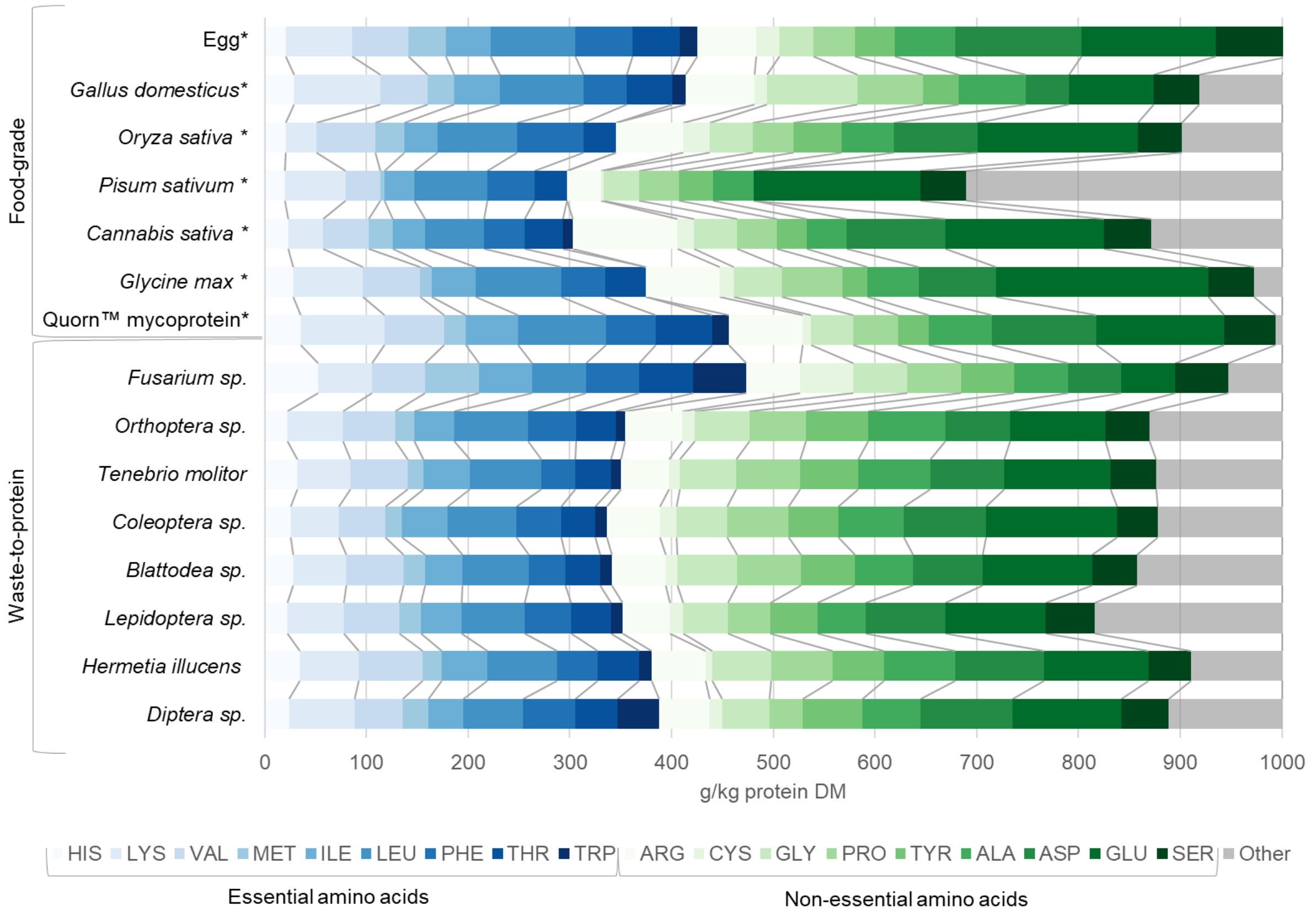

**Fig. 4. Amino acid profile of various microbial and insect protein sources.** Egg albumin is included as a standard for comparison. Eighteen amino acids are included: Histidine (HIS), Lysine (LYS), Methionine (MET), Isoleucine (ILE), Leucine (LEU), Phenylalanine (PHE), Threonine (THR), Tryptophan (TRP), Arginine (ARG), Cysteine (CYS), Glycine (GLY), Proline (PRO), Tyrosine (TYR), Alanine (ALA), Aspartic acid (ASP), Glutamic acid (GLU) and Serine (SER). We were unable to obtain values for asparagine and glutamine. Amino acid profiles are displayed for waste-to-protein protein sources including: *Fusarium sp.* (mycoprotein), *Orthoptera sp.* (crickets, grasshoppers, locusts), *Tenebrio molitor* (mealworm) *Coleoptera sp.* (beetles), *Blattodea sp.* (cockroaches, termites), *Lepidoptera sp.* (butterflies, moths), *Hermetia illucens* (black soldier fly larvae) and *Diptera sp*. Bench mark food-grade* protein sources were provided for comparison including *Gallus domesticus* (chicken), *Oryza sativa* (brown rice), *Pisum sativum* (pea), *Cannabis sativa* (hempseed), *Glycine max* (soy), and Quorn™ mycoprotein. Essential amino acid profiles are shown in blue, non-essential amino acids are shown in green on a g/kg protein dry mass basis. ‘Other’ is presented in grey and represents missing values or error due to methodology limitations reported in original literature. Detailed data can be found in Supplementary Information SI-4 and Supplementary Table ST4.

Most insects are rich in protein and other nutrients such as iron and vitamin A (*96*). Oibiopka *et al.,* (2018) found that the protein content of a diet consisting of *Orthoptera, Lepidoptera* and *Blattodea* fed to rats exhibits a 12-20% higher biological value compared to the standard protein casein (*97*). Moreover*, in vitro* digestion experiments evaluating mineral bioavailability indicated that *Orthoptera sp.* and *Tenebrio molitor* contain significantly higher chemically available calcium, magnesium, manganese, and zinc than sirloin beef (*98*).

Fig.4 shows the amino acid profiles of different food-grade benchmark animal-based, plant-based, and microbial proteins, as well as waste-to-protein insect and microbial protein sources. Compared to food-grade benchmark protein sources, waste-to-protein insect and microbial sources are richer in the essential amino acids (*99*). Waste-to-protein *Fusarium spp.* demonstrated the highest total essential amino acid contents of all protein sources, followed by food-grade egg and Quorn™ mycoprotein products, while *Diptera sp.* (including *Hermetia illucens)* protein exhibited a similar profile of essential amino acids to egg. Amongst insect proteins, *Diptera sp*. (including *Hermetia illucens)* and *Coleoptera sp.* (including *Tenebrio molitor*) appear to have the highest total amino acid contents (Fig.4). However, the nutritional quality of edible insect protein could diminish during digestion due to low content of the limiting essential amino acids, tryptophan and lysine (*40*). Previous research also reported that methionine and cysteine were limiting amino acids in *Blattodea* sp., whereas isoleucine was limiting in some *Orthoptera* sp. (*100*). Accounting for the time taken for insects to reach maturity *Hermetia illucens* and *Tenebrio molitor* larvae may be considered favourable new protein sources for rapid technology scale-up due to their relatively short lifecycles (Supplementary Information SI-6).

Depending on the grade of organic waste used as substrate, insect farming technologies provide a source of protein for human consumption or animal feed purposes. As efficient waste-to-protein bio-converters, insects achieve high conversion efficiency to turn low-grade waste into protein sources. For example, 100g of *Hermetia illucens* prepupae fed on food waste produced 80-85g of pressed cake with a high protein content of 53.1% (*101*). There is a growing number of institutions and programmes dedicated to researching insect farming as a means to address increasing global feed demands, including the International Centre of Insect Physiology and Ecology, the Sanergy project in Kenya, the Entofood partnership with Veolia in Malaysia, and Innovafeed in France (Supplementary Table ST6.4). Introducing insects such as *Hermetia illucens* as protein feed substitutes for livestock and aquaculture could bring significant socio-economic benefits such as job creation and circular economy opportunities. Overall, insects as bio-converters represents a promising technology to reintroduce waste nutrient back into the food system and shift away from land and water-dependent protein supply (*102, 103*).

**Waste-to-protein system**

A waste-to-protein system has the potential to converge waste-recovery and protein security towards a resource-circular protein future. To date, waste-to-protein technologies have been safely developed and scaled-up including the food-waste derived insect protein as animal feed (e.g. Entofood and Livalta technologies) and waste-gas to microbial protein as aquafeed (e.g. Deep Branch gas fermentation technology).

Under the waste-to-protein vision, we propose to synergistically integrate biotechnologies to maximise the recovery of food or feed-grade protein from contaminant-free organic waste while systematically considering regional characteristics on a global scale. This initiative would consider waste resource abundance and composition as well as existing industries and waste recovery infrastructure. Specifically, there is a need to develop and introduce efficient logistical approaches of supply and demand in cooperation with regulators and feed/food safety authorities.

As presented in Fig.5b, considerable amounts of food/feed grade waste are generated globally every year, including feed-grade OFMSW, lignocellulosic waste from agriculture and forestry sectors, and waste streams from the food and drink industry. Figure 5a visualises a range of chemical, physical, and biological processes that can be applied to extract protein and nutrients directly from waste, or to convert waste-carbon to sugar or other platform chemicals for subsequent protein production. Our estimated protein recovery potential was based on conversion rates (Supplementary Table ST5.3) of different technologies reported to be food- or feed-grade. With highly efficient insect bio-converters, it is estimated that 68 to 135 megatonnes/year of insect proteins could be recovered from carbon-rich OFMSW waste, depending on the insect species employed. Microbial protein technologies represent an effective lignocellulosic carbon-to-protein conversion pathway, offering protein recovery in the range of 562 megatonnes/year using food grade *F. venenatum*, or up to 1,352 megatonnes/year using feed-grade *K. marxianus* species. The estimated protein recovery potential from global food and drink industry waste (135 megatonnes/year) ranges between 15 to 22 megatonnes/year.

**a**

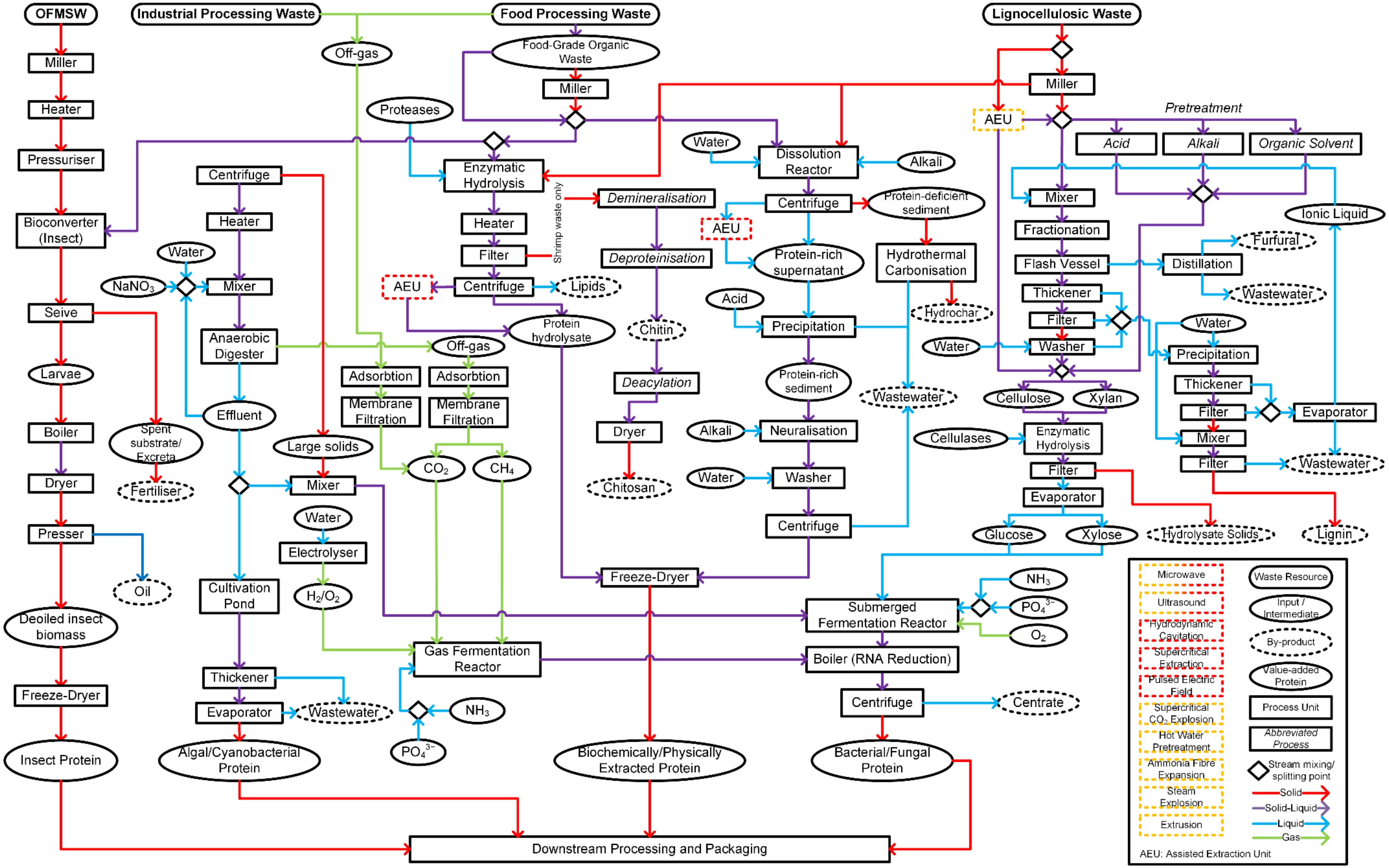

**b**

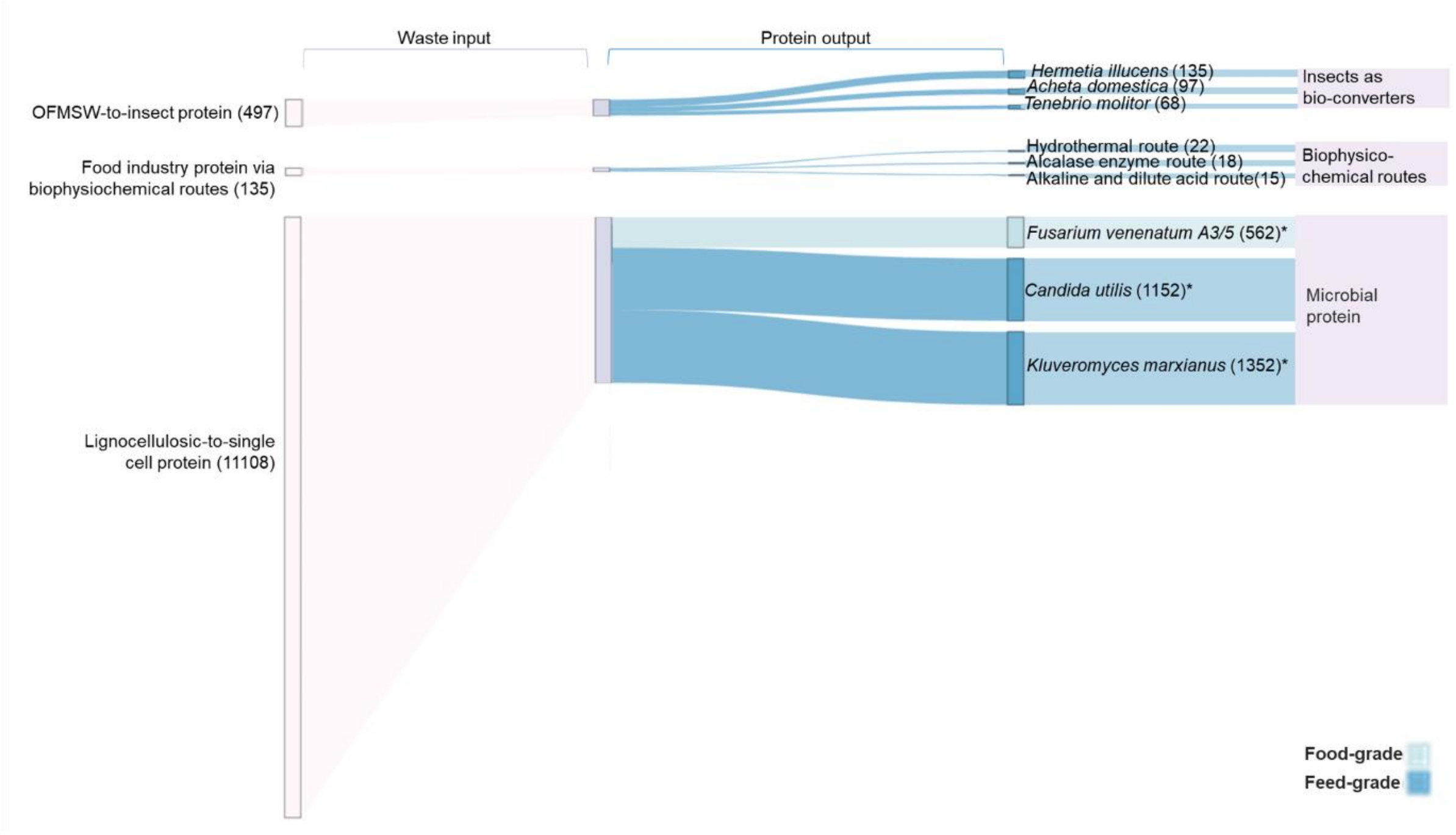

**Fig. 5. Waste-to-protein system. (a)** Process flow diagram demonstrating potential pathways for protein valorisation from organic fraction of municipal solid waste (OFMSW); industrial processing waste; food processing waste; and lignocellulosic waste to obtain value-added protein. Nodes: rectangle (rounded) = waste resource; oval (thin border) = input/intermediate; oval (dashed border) = by-product; oval (thick border) = value-added protein; rectangle = process unit; rectangle (italicised font) = abbreviated process; diamond = stream mixing/splitting point. Stream arrows: red = solid phase; purple = solid-liquid mixture; blue = liquid phase; green = gas phase. Assisted extraction unit (AEU) (red-dashed) refers to any of the following: microwave; ultrasound; supercritical extraction; pulsed electric field. AEU (orange-dashed) refers to any of the following: microwave; ultrasound; supercritical $CO_2$ explosion; hot water pretreatment; ammonia fibre expansion; steam explosion; extrusion (*45, 104-117*). **(b)** Quantitative mass balance for a theoretical waste-to-protein system. Input waste streams are shown on the left: OFMSW-to-insect protein, agricultural lignocellulosic-to-microbial protein, and food industry (including fishing, aquaculture and brewery industry) protein via biophysiochemical routes. The protein outputs are shown on the right. **Candida Utilis* and *Kluveromyces marxianus* are capable of utilising hexose and pentose sugars. Values given are for glucose utilisation only. Inclusion of pentose sugars increases conversion outputs to an upper range

of 893 megatonne/year, 1,831 megatonne/year and 2,149 megatonne/year for *Fusarium venenatum A3/5*, *Candida utlilis,* and *Kluveromyces marxianus,* respectively. All values in brackets are given in megatonne/year. Detailed data can be found in Supplementary Information SI-5 and Supplementary Table ST5.

However, our estimated recovery value focuses on *F. venenatum* due to its history as an widely-accepted food source (*73*). This pathway offers a potential 562 megatonnes/year recovery of food protein from the 11,108 megatonnes/year cellulosic waste content produced by global agricultural and forestry sectors, supplying 72g/capita/year (197g/capita/day) waste-derived protein to meet the average adult daily protein recommendation (50g per 70kg) (*118*). However, as these estimates were based on conversion rates derived from literature data, further characterisation of region-specific waste composition and exploratory research on resource recovery potential at scale are essential to provide evidence for informed decision-making.

It should be noted that both waste compositions (Fig.1 and Fig.2) and existing waste recovery systems differ significantly across countries. Developed and urbanised regions tend to produce higher quantities of MSW with a lower organic component than low-income countries and offer established centralised waste collection and treatment infrastructure. Thus, centralised waste-to-protein systems represent great potential for increased efficiency (*15*). In less developed countries, there are still large amounts of untapped waste resources including OFMSW, agricultural and forestry lignocellulosic waste, and food and drink industry waste that represent unexploited future potential for a waste-to-protein system (*25*). The more sporadic distribution of organic waste and lack of sustainable waste-recovery systems positions decentralised waste-to-protein solutions as the most suitable approach for such countries. Examples include those in recent studies focused on *Hermetia illucens* as bio-converters of food processing waste, and microbial protein routes developed by Deep Branch for aquafeed production from decentralised waste gas streams (*119*, *120*). The significant global variations discussed call for a systems approach to synergistically integrate centralised and decentralised technologies and optimise waste-to-protein solutions which consider the spatial distribution of regional waste and existing industries and infrastructures.

Perceptions of a 'waste-to-protein' concept vary significantly by country and also warrant consideration. African and South East Asian countries appear to be good candidates for expansion of technologies that utilise insects as bio-converters due to their relatively strong cultural acceptance of insects as food (*121*). Microbial fermentation is already well-established in Europe and North America, with Quorn™ being a popular and mainstream food product in both regions**.** These regions would therefore be a good target for expanding microbial protein technologies. It is essential that upgraded 'waste-to-protein' products are regarded as high-quality and safe by consumers globally. As such, conversion and upgrading must proceed within the conditions set out by the feed/food chain alliance and must comply with hygiene quality and safety standards set by regulators (*122*, *123*).

New protein sources have been highlighted as novel food, which need to meet general food safety requirements stipulated in national or regional food safety regulations (*124*). Global approaches to the regulation of novel food vary significantly. In the EU, Canada, Singapore, and India, evidence of 'history of safe use' (HOSU) is considered globally, whereas in China, Australia/New Zealand (AU/NZ) and Brazil, the scope of HOSU is restricted to native consumption (*125*-*129*). AU/NZ and Canada are exceptional in that there is no rigid cut-off date defined for HOSU, giving their respective regulatory authorities an extra degree of freedom in determining novel status (*125*). In these countries, if a protein for food purposes is deemed novel by the responsible authoritative body, a risk assessment is then undertaken considering evidence submitted in the form of a dossier by the manufacturer (*125*-*129*). Pre-submission consultations can help to identify missing information and errors in the dossier to avoid ‘clock-stop’ delays in the risk assessment stage. Food Standards Australia/New Zealand, Singapore Food Association and Health Canada have established organisations specifically for this purpose (*125, 126*). In the US, novel status is commonly self-determined by the manufacturer in accordance with generally recognised as safe (GRAS) standards, through convening of an expert panel to review publicly available scientific data on the HOSU of their product (*125*). Alternatively, a food additive petition can be submitted to the Food and Drug Administration. However, data from in-house testing pertaining to safety of the product is required in this case, incurring similar issues of high cost and extended timelines from submission-to-market as in countries adopting an EU-style model (*125, 130*). Further details on global novel food/protein regulations and notification processes can be found in Supplementary Information SI-7. Recent regulatory advances on waste-to-protein for animal feed purposes in the EU includes Regulation (EU) 2021/1372, an amendment that allows the use of insect-processed proteins as feed (*131*). Subsequently, Regulation (EU) 2021/1925 was implemented to authorise the use of *Bombyx mori* (silkworm) processed animal proteins in animal feed, the eighth insect species to be approved (*132*).

Insect proteins and microbial proteins offer environmental advantages over conventional animal-source or plant-sourced proteins, in particular on climate change mitigation and arable land use reduction (Supplementary Information SI-8 and Supplementary Table ST7). However, novel protein research and technologies are still at the infant stage in contrast to conventional protein sources, which operate at higher technological readiness levels (TRL) 7-9. Thus, future research into waste-to-protein scale-up potential, particularly with regards to process integration and optimisation, is necessary to enable novel waste-to-protein products to become economically competitive. Nevertheless, new protein sources have the potential to contribute towards food systems that operate within scientifically defined targets for sustainability, both at local and Earth system scales, i.e. planetary boundaries (*133*).

Overall, it is not only conversion efficiency and nutritional quality of proteins recovered from waste that are of importance, but also the processability, scalability and acceptability of a waste-to-protein system that are highly relevant to future work. Thus, future research and technology development should focus on the waste resources and protein solutions that i) offer food- or feed-grade nutrition values; ii) are easily processed and harvested, and thereby able to fit into existing food supply chains; and iii) consider perception, safety and acceptability to the consumers and regulators; and iv) advance the understanding of waste-to-protein technology performance, including process optimisation at scale, techno-economic viability, and environmental sustainability.

## Conclusions

Animal-sourced proteins are not only carbon-intensive and resource-demanding, but also vulnerable to pandemic effects (e.g. Covid-19) due to long-production cycles (except for chicken, Supplementary information SI-6) and animals being susceptible to infection. These factors, combined with increasing protein demands and the persistent global hunger pandemic, highlight the complex challenges of ensuring protein security for human health within environmental boundaries. In this quantitative analysis, we proposed a waste-to-protein upgrading system. By synergistically integrating waste-to-protein technologies, this system has the potential to solve a significant component of the global challenge of a planet degrading food system and converge innovations on zero-waste and protein security towards a sustainable protein future. Our study emphasises the importance of upstream quality preservation by assuring contaminant-free organic waste streams and systems analysis to estimate the waste-to-protein potential involving chemical, physical, and biological conversion pathways. We quantified global waste streams, which are rich in carbon and nutrients and absent of pathogens and hazardous substances. These streams present a global annual resource potential of 497 megatonnes of OFMSW, 135 megatonnes of by-products from the brewing and shrimp fishing industries and 11108 megatonnes of lignocellulosic agricultural and forestry waste (equivalent to 9386 megatonnes of holocellulosic contents, which can be converted to fermentable sugars amounting to 2503 megatonnes of glucose, or 3980 megatonnes of glucose and xylose). Over 80 microbial species have been discovered to enable efficient waste recovery of microbial protein with preferable amino acid profiles that are characteristic of proteins of high biological value. A concerted effort to broaden the range of micro-organisms is warranted, either independently or in combination with microbiomes or designed cultures that can be regarded as safe for upgrading secondary resources to safe feed and food. Insects as bio-converters offer efficient mechanisms to convert different grades of waste to food or feed proteins, which are generally rich in protein, vitamins, and minerals such as iron, calcium, manganese and zinc compared with other animal-sourced proteins.

Despite advances in individual technologies, critical gaps remain in the development of innovative systems which will enable ‘plug-and-play’ solutions, synergistic technology integration, and optimisation of the protein recovery from diverse waste streams. Although we demonstrate that waste-to-protein system has the potential to recover waste and catalyse novel protein solutions, scientific targets that define healthy and sustainable protein production remain absent. Integrated assessment and optimisation of waste-to-protein value chains that consider scientifically quantified planetary boundaries (*133*) represent a future research frontier to further understand the implications of a waste-to-protein transition for water, land, biodiversity, carbon, nitrogen and phosphorus (5 of the 9 planetary boundaries). Notably, evidence-informed regulatory response timelines are considerably lagging behind the accelerated food and feed technology innovations including novel proteins. For waste-to-protein, many aspects remain unknown, such as the quality of low-value waste streams, nutritional values and health effects. Such regulatory barriers hinder the development of waste-to-protein technologies. Future research to enable deep scanning of the fast-paced protein innovation landscape and develop a system for rapid regulatory response is needed. A sustainable protein system can only be achieved by multi-sector, multi-level actions that include a substantial global shift towards reduction in food loss and waste, and deployment of innovative protein technologies. Under the international policy framework, human health and environmental sustainability are included in most of the United Nations Sustainable Development Goals (SDGs) (*134*). Integrated analyses of different future diet and protein scenarios and their impacts on humans (SDGs 1 and 2) and on planetary boundaries (SDGs 6, 13, 14, 15, on water, climate, ocean, and biodiversity) are necessary to inform future policy and technology development. A crucial element is the linkage of the waste-to-protein supply chains, environment footprint and the overall regulatory measures in relation to the sustainability and safety of upgrade-protein to help ameliorate the persistent and ongoing hunger pandemic and to protect the planet.

## Acknowledgement

**Funding:** G.K., A.K., H.F.R. and M.G. would like to acknowledge the UK Royal Academy of Engineering for Frontiers of Engineering for Development Seed funding. E.P. and M.G. would like to acknowledge the UK Engineering and Physical Sciences Research Council (EPSRC) and Mond Nissin Corporate for providing financial support under EPSRC iCASE programme. **Author contributions:** M.G., G.K., A.K., H.F.R. designed the study. E.P., M.B., Y.H. and M.G performed research and data analyses. E.P., W.V., P.E., and M.G., drafted the manuscript sections. M.B. provided additional text and edits. J.R., P.S., O.W., C.H., J.H., contributed significantly to the paper revision. A.L. and M.B. contributed to the data visualisation. All authors contributed to the final paper revision and approved the final manuscript. **Competing interests:** The authors declare no competing interests. **Data and materials availability:** All data needed to evaluate the conclusions in the paper are present in the paper, Supplementary Information, Extended Tables, and/or in the materials cited herein. Correspondence and requests for further materials should be addressed to Miao Guo (miao.guo@imperial.ac.uk). **Other:** E.P and M.G would like to acknowledge Thomas Upcraft for providing data on lignocellulosic mycoprotein technologies.